\documentstyle[twoside,fleqn,espcrc2,epsf]{article}

\def\gsim{\mathrel{\raise2pt\hbox to 8pt{\raise -5pt\hbox{$\sim$}\hss{$>$}}}}
\def\rsim{\mathrel{\raise2pt\hbox to 8pt{\raise -5pt\hbox{$\sim$}\hss{$>$}}}}
\def\lsim{\mathrel{\raise2pt\hbox to 8pt{\raise -5pt\hbox{$\sim$}\hss{$<$}}}}

\title{Topological Density and Instantons on a Lattice
\thanks{Presented by J.~Grandy.  Computer
time on C90 was provided by NERSC and PSC Centers.}}

\author{J.~Grandy\address{Physics Department, 174 West 18th Avenue, 
Ohio State University, Columbus, OH 43210}
        and R.~Gupta\address{T-8 Group, MS B285, Los Alamos National
Laboratory, Los Alamos, New Mexico 87545 U.~S.~A.~}}

\begin{document}

\begin{abstract}
Abstract: We present an update on the study of topological structure
of QCD.  Issues addressed include a comparison between the plaquette
and the geometric methods of calculating the topological density.  We
show that the improved gauge action based on $\sqrt3$ blocking
transformation suppresses the formation of topologically charged
dislocations with low action.  Using a cooling method we identify the
instantons' location, estimate their size and density, and calculate the
renormalization constant $Z_Q$ for the plaquette method.

\end{abstract}

\maketitle

\setlength{\textfloatsep}{12pt plus 2pt minus 2pt}

Issues related to the discretization of the 
continuum expression for the topological density,
\begin{equation} 
q(x) = {{-1}\over{32\pi^2}} {\rm Tr}\left(\epsilon_{\mu\nu\rho\sigma}
F_{\mu\nu}F_{\rho\sigma}\right) 
\end{equation} 
depend on which of several methods, of which we consider the
plaquette\cite{INFN90}, and geometric (L\"uscher's \cite{Luscher}), is
used to calculate $q(x)$, and on the lattice action.  The geometric
method while preserving an integer result for the topological charge
${Q=\sum_x q(x)}$
gives rise to a divergent topological susceptibility $\chi_t$ in the
continuum limit due to lattice artifacts called
dislocations\cite{Gockeler89}.  We show that these dislocations are
much better controlled by an improved action (IA) derived by
approximating the renormalization trajectory(RT) of the ($\sqrt{3}$)
block transformation \cite{ImprAct}.  We also show that
cooling\cite{CoolRefs} is useful for quantifying instantons' locations
and sizes, and for estimating the large multiplicative renormalization
$Z_Q(\beta=6)$ for the plaquette method\cite{INFN90,DiGiacomo94}.

Our calculations are performed on the existing ensemble of $34\  (16^3\times40)$
lattices at $\beta=6.0$ using the Wilson action \cite{lat93}.  For the RGT improved
action, we use $55\ (18^3\times36)$ lattices at $\beta \approx 6.0$. 

\section{Improved Action: Geometric Method}

On the lattice the action of an instanton goes to zero as the core
size $\rho \to 0$, rather than show the expected logarithmic
divergence.  The charge, measured using the geometric method, jumps
from $0$ to $1$ at some $\rho_c$ where the lattice action $S_{min}$ is
still much smaller than that for the classical instanton.  These
boundary configurations with $Q = \pm 1$ are called dislocations, and
are lattice artifacts because their size $\rho < a$ for all $g$,
$i.e.$ $\rho_c$ does not scale as expected.  In the case of the Wilson
action (WA) the Boltzmann suppression $e^{-\beta S_{min}}$ is
overwhelmed by the entropy factor $e^{\beta 16\pi^2/33}$, and
dislocations dominate the calculation of $\chi_t$ in the continuum
limit.  To raise $S_{min}$ we investigate an IA consisting of the
plaquette in the fundamental, $8$ and $6$ representations and the
$1\times 2$ planar loop with couplings \cite{ImprAct} \looseness -1
\begin{equation} 
{K_{1\times 2} \over K_F} = -0.04; 
{K_{8} \over K_F} = -0.12; 
{K_{6} \over K_F} = -0.12.
\label{eq:action}
\end{equation} 
In our normalization, the bare coupling is given by $K_F =
c_{(1\times1)} 6 /g^2 $ where $c_{(1\times1)}$ is the weight
associated with the plaquette.  We estimate $S_{min}^{IA}$ by the
following method.  We constrain a central plaquette to be $-1$ in a
$SU(2)$ projection, a characteristic of most dislocations that have
been considered, and then vary the links around the plaquette in a
collective fashion until a minimum of the action is found with a
topological charge of $1$.  For such configurations
$S_{min}^{IA} = 18.9$, even larger than the value $8\pi^2/6$
for a classical instanton. A comparison of $S_{min}^{IA}$ with other
actions studied in \cite{Gockeler89} is made in Table \ref{tab:smin},
along with $c_{(1\times1)}$ for each action.  We conjecture that
dislocations are suppressed for improved actions with large
$c_{(1\times1)}$ because the central plaquette near the cut locus
contributes a significant fraction to the total action of the
dislocation and a large $c_{(1\times1)}$ raises the action of this
central plaquette.

\begin{table}[t]
\setlength{\tabcolsep}{7.7pt}
\caption{Minimum action of configurations with $Q=1$}
\begin{tabular}{|r|c|r|l|}
\hline
\multicolumn{1}{|c|}{Action} & 
\multicolumn{1}{|c|}{$c_{(1\times1)}$} & 
\multicolumn{1}{|c|}{SU(3) $\bar{S}_{min}$} &
\multicolumn{1}{|c|}{Ref.} \cr \hline
Wilson & $1.0$ & $4.5$ \hspace{15pt} & \cite{Gockeler89} \cr
Symanzik & $1.67$ & $6.1$  \hspace{15pt} & \cite{Gockeler89} \cr
Wilson Impr. & $4.376$ & $10.7$ \hspace{15pt} & \cite{Gockeler89} \cr
$\sqrt{3}$ RGT & $9.09$ & $18.9$ \hspace{15pt} &   \cr \hline
\end{tabular}
\label{tab:smin}
\end{table}

The results for $\chi_t$ by the geometric method at $K_F=10.58$ and
for the WA \cite{lat93} at $\beta=6.0$ are given in
Table~\ref{tab:chit}. A rough estimate of the scale for the IA is
$a^{-1} = 2.3\,{\rm GeV}$, $i.e.$ the same as for the WA at $\beta=
6.0$.  $\chi_t$ with the WA lies significantly above that predicted by
the Witten-Veneziano formula\cite{Veneziano}, however, due to the
uncertainty in scale for the IA we cannot yet isolate the role of
dislocations by comparing Wilson and improved action results or give a
reliable estimate for $\chi_t^{IA}$.  However, the two advantages of
the IA in the study of topology are (i) raising $S_{min}$ to suppress
dislocations and (ii) the IA plaquette expectation value $\langle \Box
\rangle = 1.871$, is much higher than the WA value $1.782$,
consequently the integration routines for calculating $q(x)$ converge
about $70\%$ faster as is expected of smoother fields.

\begin{table}[b]
\setlength{\tabcolsep}{4.8pt}
\caption{Topological susceptibility by geometric method}
\begin{tabular}{|r|c|c|l|}
\hline
\multicolumn{1}{|c|}{Action} & 
\multicolumn{1}{|c|}{$\chi_t^{1/4}a$} & 
\multicolumn{1}{|c|}{$\chi_t$ (MeV) } &
\multicolumn{1}{|c|}{Ref.} \cr \hline
Wilson & $0.114(8)$ & $262(18)$  & \cite{lat93} \cr
$\sqrt{3}$ RGT & 0.0975(46) & $224(11)$ & (this work) \cr \hline
\end{tabular}
\label{tab:chit}
\end{table}

\section{Identifying instantons via Cooling}

In order to correlate hadronic structure with topology we need to
identify the shapes and sizes of all instantons on the uncooled
lattices.  Ultraviolet (UV) noise in $q(x)$ on the uncooled lattices
has thwarted our efforts to devise a method to do this.  We therefore
adiabatically\cite{longpaper} cool a {\it subset} of lattices (32 WA
lattices and 28 IA lattices) for a total cooling time $\tau=10$ and
measure $q(x)$ using the plaquette and L\"uscher's method for cooling
times $\tau = 0, 0.3, 1, 3, 6, 10$. The results are shown in
Table~\ref{tab:chicool}, using $Z_Q(\tau)$ as estimated below.  Note
that for $\tau \geq 1$ the two methods give consistent results.  The
complete lack of dependence of $\chi_t^{geom}$ on $\tau$ is probably
fortuitous.

On the $\tau=10$ cooled lattices we identify hyperspherical volumes of
radius $r^2\le r_{max}^2$ containing a total net geometric charge of
magnitude greater than $q_{min}$ (we generally set $q_{min} =1$) and
consider these volumes to contain an instanton or anti-instanton.  Our
search routine first isolates small instantons and excludes their
volumes from future search.  As a result, if there are overlapping
instantons, larger instantons ($r_{max}^2 \rsim 30$) lose the
spherical shape we normally associate with classical instantons,
therefore for each instanton volume containing $v$ points we define an
effective radius $r_{eff} = (2v/\pi^2)^{1/4}$.  Figure
\ref{fig:reffch} shows the distribution of instanton volumes for IA in
terms of $r_{eff}^2$, with $r_{max}^2=35$.  Most are in the range
$4a\leq r_{eff}\leq 5.5a$, or about $0.35$ to $0.5\ {\rm fm}$.  Taking
the mean size to be $r_{eff} = 5$ and assuming a uniform distribution,
we estimate the core size of classical instantons to be $\rho \sim
0.3\ {\rm fm}$. On WA lattices we find a total of $N_+ = 180$
instantons and $N_-=200$ anti-instantons, corresponding to an
instanton density of $1.3 {\rm fm}^{-4}$.  Similarly, for the IA we
find $N_+ = 440$, $N_-=421$, which gives a factor of two larger
density of $2.7\ {\rm fm}^{-4}$.

\ifx\axisloaded\relax \fi






 
\def\setRevDate $#1 #2 #3${\def\TeXdrawId{TeXdraw V1R3 revised <#2>}}
\setRevDate $Date: Wed Mar  3 12:01:12 MST 1993$

\chardef\catamp=\the\catcode`\@
\catcode`\@=11
\ifx\TeXdraw@included\undefined\global\let\TeXdraw@included=\relax\else
\errhelp{TeXdraw needs to be input only once outside of any groups.}%
\errmessage{Multiple call to include TeXdraw ignored}%
\expandafter \fi

\long                              
\def\centertexdraw #1{\hbox to \hsize{\hss
                                      \btexdraw #1\etexdraw
                                      \hss}}


\def\btexdraw {\x@pix=0             \y@pix=0
               \x@segoffpix=\x@pix  \y@segoffpix=\y@pix
               \t@exdrawdef
               \setbox\t@xdbox=\vbox\bgroup\offinterlineskip
                   \global\d@bs=0           
                   \t@extonlytrue           
                   \p@osinitfalse
                   \savemove \x@pix \y@pix  
                   \m@pendingfalse
                   \p@osinitfalse           
                   \p@athfalse}


\def\etexdraw {\ift@extonly \else
                 \t@drclose      
               \fi
               \egroup           
               \ifdim \wd\t@xdbox>0pt
                 \errmessage{TeXdraw box non-zero size,
                             possible extraneous text}%
               \fi
               \maxhvpos         
               \pixtodim \xminpix \l@lxpos  \pixtodim \yminpix \l@lypos
               \pixtobp {-\xminpix}\l@lxbp  \pixtobp {-\yminpix}\l@lybp
               \vbox {
		      \offinterlineskip
                      \ift@extonly \else
                        \includepsfile{\p@sfile}{\the\l@lxbp}{\the\l@lybp}%
				      {\the\hdrawsize}{\the\vdrawsize}%
                      \fi
                      \vskip\vdrawsize
                      \vskip \l@lypos
                      \hbox {\hskip -\l@lxpos
                             \box\t@xdbox
                             \hskip \hdrawsize
                             \hskip \l@lxpos}%
                      \vskip -\l@lypos\relax}}

%

\def\drawdim #1 {\def\d@dim{#1\relax}}


\def\setunitscale #1 {\edef\u@nitsc{#1}%
                      \realmult \u@nitsc  \s@egsc \d@sc}
\def\relunitscale #1 {\realmult {#1}\u@nitsc \u@nitsc
                      \realmult \u@nitsc \s@egsc \d@sc}
\def\setsegscale #1 {\edef\s@egsc {#1}%
                     \realmult \u@nitsc \s@egsc \d@sc}
\def\relsegscale #1 {\realmult {#1}\s@egsc \s@egsc
                     \realmult \u@nitsc \s@egsc \d@sc}

\def\bsegment {\ifp@ath
                 \flushmove
               \fi
               \begingroup
               \x@segoffpix=\x@pix
               \y@segoffpix=\y@pix
               \setsegscale 1
               \global\advance \d@bs by 1 }
\def\esegment {\endgroup
               \ifnum \d@bs=0
                 \writetx {es}%
               \else
                 \global\advance \d@bs by -1
               \fi}

\def\savecurrpos (#1 #2){\getsympos (#1 #2)\a@rgx\a@rgy
                         \s@etcsn \a@rgx {\the\x@pix}%
                         \s@etcsn \a@rgy {\the\y@pix}}%
\def\savepos (#1 #2)(#3 #4){\getpos (#1 #2)\a@rgx\a@rgy
                            \coordtopix \a@rgx \t@pixa
                            \advance \t@pixa by \x@segoffpix
                            \coordtopix \a@rgy \t@pixb
                            \advance \t@pixb by \y@segoffpix
                            \getsympos (#3 #4)\a@rgx\a@rgy
                            \s@etcsn \a@rgx {\the\t@pixa}%
                            \s@etcsn \a@rgy {\the\t@pixb}}

\def\linewd #1 {\coordtopix {#1}\t@pixa
                \flushbs
                \writetx {\the\t@pixa\space sl}}
\def\setgray #1 {\flushbs
                 \writetx {#1 sg}}
\def\lpatt (#1){\listtopix (#1)\p@ixlist
                \flushbs
                \writetx {[\p@ixlist] sd}}

\def\lvec (#1 #2){\getpos (#1 #2)\a@rgx\a@rgy
                  \s@etpospix \a@rgx \a@rgy
                  \writeps {\the\x@pix\space \the\y@pix\space lv}}
\def\rlvec (#1 #2){\getpos (#1 #2)\a@rgx\a@rgy
                   \r@elpospix \a@rgx \a@rgy
                   \writeps {\the\x@pix\space \the\y@pix\space lv}}
\def\move (#1 #2){\getpos (#1 #2)\a@rgx\a@rgy
                  \s@etpospix \a@rgx \a@rgy
                  \savemove \x@pix \y@pix}
\def\rmove (#1 #2){\getpos (#1 #2)\a@rgx\a@rgy
                   \r@elpospix \a@rgx \a@rgy
                   \savemove \x@pix \y@pix}

\def\lcir r:#1 {\coordtopix {#1}\t@pixa
                \writeps {\the\t@pixa\space cr}%
                \r@elupd \t@pixa \t@pixa
                \r@elupd {-\t@pixa}{-\t@pixa}}
\def\fcir f:#1 r:#2 {\coordtopix {#2}\t@pixa
                     \writeps {#1 \the\t@pixa\space fc}%
                     \r@elupd \t@pixa \t@pixa
                     \r@elupd {-\t@pixa}{-\t@pixa}}
\def\lellip rx:#1 ry:#2 {\coordtopix {#1}\t@pixa
                         \coordtopix {#2}\t@pixb
                         \writeps {\the\t@pixa\space \the\t@pixb\space el}%
                         \r@elupd \t@pixa \t@pixb
                         \r@elupd {-\t@pixa}{-\t@pixb}}
\def\larc r:#1 sd:#2 ed:#3 {\coordtopix {#1}\t@pixa
                            \writeps {\the\t@pixa\space #2 #3 ar}}


\def\ifill f:#1 {\writeps {#1 fl}}     
\def\lfill f:#1 {\writeps {#1 fp}}     



\def\htext #1{\def\testit {#1}%
              \ifx \testit\l@paren
                \let\next=\h@move
              \else
                \let\next=\h@text
              \fi
              \next{#1}}

\def\rtext td:#1 #2{\def\testit {#2}%
                    \ifx \testit\l@paren
                      \let\next=\r@move
                    \else
                      \let\next=\r@text
                    \fi
                    \next td:#1 {#2}}

\def\vtext {\rtext td:90 }

\def\textref h:#1 v:#2 {\ifx #1R%
                          \edef\l@stuff {\hss}\edef\r@stuff {}%
                        \else
                          \ifx #1C%
                            \edef\l@stuff {\hss}\edef\r@stuff {\hss}%
                          \else  
                            \edef\l@stuff {}\edef\r@stuff {\hss}%
                          \fi
                        \fi
                        \ifx #2T%
                          \edef\t@stuff {}\edef\b@stuff {\vss}%
                        \else
                          \ifx #2C%
                            \edef\t@stuff {\vss}\edef\b@stuff {\vss}%
                          \else  
                            \edef\t@stuff {\vss}\edef\b@stuff {}%
                          \fi
                        \fi}

\def\avec (#1 #2){\getpos (#1 #2)\a@rgx\a@rgy
                  \s@etpospix \a@rgx \a@rgy
                  \writeps {\the\x@pix\space \the\y@pix\space (\a@type)
                            \the\a@lenpix\space \the\a@widpix\space av}}

\def\ravec (#1 #2){\getpos (#1 #2)\a@rgx\a@rgy
                   \r@elpospix \a@rgx \a@rgy
                   \writeps {\the\x@pix\space \the\y@pix\space (\a@type)
                             \the\a@lenpix\space \the\a@widpix\space av}}

\def\arrowheadsize l:#1 w:#2 {\coordtopix{#1}\a@lenpix
                              \coordtopix{#2}\a@widpix}
\def\arrowheadtype t:#1 {\edef\a@type{#1}}

\def\clvec (#1 #2)(#3 #4)(#5 #6)%
           {\getpos (#1 #2)\a@rgx\a@rgy
            \coordtopix \a@rgx\t@pixa
            \advance \t@pixa by \x@segoffpix
            \coordtopix \a@rgy\t@pixb
            \advance \t@pixb by \y@segoffpix
            \getpos (#3 #4)\a@rgx\a@rgy
            \coordtopix \a@rgx\t@pixc
            \advance \t@pixc by \x@segoffpix
            \coordtopix \a@rgy\t@pixd
            \advance \t@pixd by \y@segoffpix
            \getpos (#5 #6)\a@rgx\a@rgy
            \s@etpospix \a@rgx \a@rgy
            \writeps {\the\t@pixa\space \the\t@pixb\space 
                      \the\t@pixc\space \the\t@pixd\space 
                      \the\x@pix\space \the\y@pix\space cv}}
\def\e@tendspline#1\endpoints{}
\newtoks\splinet@ks
\def\splinep@int #1 #2 %
           {%
            \advance\p@intnumber by 1\relax
	    \splinet@ks={(#1 #2)}%
            \us@rconvert (#1 #2)\a@rgx\a@rgy
            \futurelet\n@xttok\wr@tesplinepoint}
\def\wr@tesplinepoint{%
            \ifx\n@xttok\endpoints
              \s@etpospix \a@rgx \a@rgy
	      \expandafter
	      \us@rfinish \expandafter
              \p@intnumber \expandafter:\the\splinet@ks=({\x@pix} {\y@pix})%
              \expandafter\e@tendspline
            \else
              \coordtopix \a@rgx\t@pixa
              \advance \t@pixa by \x@segoffpix
              \coordtopix \a@rgy\t@pixb
              \advance \t@pixb by \y@segoffpix
	      \expandafter
              \us@rpoint \expandafter
              \p@intnumber \expandafter:\the\splinet@ks=({\t@pixa} {\t@pixb})%
              \expandafter\splinep@int
            \fi}
\def\defaultus@rfinish#1:(#2 #3)=(#4 #5){\writeps {\the#4 \the#5 \the#1 BSpl}}
\def\defaultus@rpoint#1:(#2 #3)=(#4 #5){\writeps {\the#4 \the#5}}
\def\s@tuseroptions#1/#2/#3/#4@{%
\let\us@rconvert=#1\relax\ifx\us@rconvert\relax\let\us@rconvert=\getpos\fi
\let\us@rpoint=#2\relax\ifx\us@rpoint\relax\let\us@rpoint=\defaultus@rpoint\fi
\let\us@rfinish=#3\relax\ifx\us@rfinish\relax\let\us@rfinish=\defaultus@rfinish
   \fi}%
\def\dooverpoints#1\points{%
    \p@intnumber=0\relax\s@tuseroptions#1///@\splinep@int}
\def\spline{\dooverpoints\points}

\newcount\p@intnumber
\def\curvytype#1{\def\curv@type{#1}}\curvytype{4}%
\def\curvyheight#1{\def\curv@height{#1}}\curvyheight{10}%
\def\curvylength#1{\def\curv@length{#1}}\curvylength{10}%
\def\drawcurvyphoton around (#1 #2) from (#3 #4) to (#5 #6)%
 {\getpos (#1 #2)\a@rgx\a@rgy
  \coordtopix \a@rgx \t@pixa \advance \t@pixa by \x@segoffpix
  \coordtopix \a@rgy \t@pixb \advance \t@pixb by \y@segoffpix
  \writeps {mark \the\t@pixa\space \the\t@pixb}%
  \getpos (#3 #4)\a@rgx\a@rgy
  \s@etpospix \a@rgx\a@rgy
  \writeps {\the\x@pix\space \the\y@pix}%
  \getpos (#5 #6)\a@rgx\a@rgy
  \s@etpospix \a@rgx\a@rgy
  \writeps {\the\x@pix\space \the\y@pix}%
  \writeps {\curv@height\space \curv@length\space \curv@type\space%
              curvyphoton}%
}%
\def\drawcurvygluon around (#1 #2) from (#3 #4) to (#5 #6)%
 {\getpos (#1 #2)\a@rgx\a@rgy
  \coordtopix \a@rgx \t@pixa \advance \t@pixa by \x@segoffpix
  \coordtopix \a@rgy \t@pixb \advance \t@pixb by \y@segoffpix
  \writeps {mark \the\t@pixa\space \the\t@pixb}%
  \getpos (#3 #4)\a@rgx\a@rgy
  \s@etpospix \a@rgx\a@rgy
  \writeps {\the\x@pix\space \the\y@pix}%
  \getpos (#5 #6)\a@rgx\a@rgy
  \s@etpospix \a@rgx\a@rgy
  \writeps {\the\x@pix\space \the\y@pix}%
  \writeps {\curv@height\space 2 mul \curv@length\space \curv@type\space%
              curvygluon}%
}%
\def\blobfreq#1{\def\bl@bfreq{#1}}\blobfreq{0.2}%
\def\blobangle#1{\def\bl@bangle{#1}}\blobangle{0}%
\def\hatchedblob#1{\def\bl@btype{(#1)}}\hatchedblob{B}%
\def\grayblob#1{\def\bl@btype{#1}}%
\def\drawblob xsize:#1 ysize:#2 at (#3 #4)%
{\getpos (#3 #4)\a@rgx\a@rgy \s@etpospix \a@rgx\a@rgy
 \writeps{\the\x@pix\space \the\y@pix}%
 \getpos (#1 #2)\a@rgx\a@rgy 
 \coordtopix \a@rgx\t@pixa \coordtopix\a@rgy\t@pixb \writeps
{\the\t@pixa\space\the\t@pixb\space\bl@bangle\space\bl@bfreq\space\bl@btype}%
 \writeps {blob}%
 \rmove (-#1 -#2)\rmove (#1 #2)\rmove (#1 #2)\rmove (-#1 -#2)}%
\def\drawbb {\bsegment
               \drawdim bp
               \setunitscale 0.24
               \linewd 1           
               \writeps {\the\xminpix\space \the\yminpix\space mv}%
               \writeps {\the\xminpix\space \the\ymaxpix\space lv}%
               \writeps {\the\xmaxpix\space \the\ymaxpix\space lv}%
               \writeps {\the\xmaxpix\space \the\yminpix\space lv}%
               \writeps {\the\xminpix\space \the\yminpix\space lv}%
             \esegment}


\def\getpos (#1 #2)#3#4{\g@etargxy #1 #2 {} \\#3#4%
                        \c@heckast #3%
                        \ifa@st
                          \g@etsympix #3\t@pixa
                          \advance \t@pixa by -\x@segoffpix
                          \pixtocoord \t@pixa #3
                        \fi
                        \c@heckast #4%
                        \ifa@st
                          \g@etsympix #4\t@pixa
                          \advance \t@pixa by -\y@segoffpix
                          \pixtocoord \t@pixa #4
                        \fi}

\def\getsympos (#1 #2)#3#4{\g@etargxy #1 #2 {} \\#3#4%
                           \c@heckast #3%
                           \ifa@st \else
                             \errmessage {TeXdraw: invalid symbolic coordinate}
                           \fi
                           \c@heckast #4%
                           \ifa@st \else
                             \errmessage {TeXdraw: invalid symbolic coordinate}
                           \fi}

\def\listtopix (#1)#2{\def #2{}%
                      \edef\l@ist {#1 }
                      \t@countc=0
                      \loop
                        \expandafter\g@etitem \l@ist \\\a@rgx\l@ist
                        \a@pppix \a@rgx #2
                        \ifx \l@ist\empty
                          \t@countc=1
                        \fi
                      \ifnum \t@countc=0
                      \repeat}


\def\realmult #1#2#3{\dimen0=#1pt
                     \dimen2=#2\dimen0
                     \edef #3{\expandafter\c@lean\the\dimen2}}

\def\intdiv #1#2#3{\t@counta=#1
                   \t@countb=#2
	           \ifnum \t@countb<0
                      \t@counta=-\t@counta
                      \t@countb=-\t@countb
                   \fi
                   \t@countd=1                    
                   \ifnum \t@counta<0
                      \t@counta=-\t@counta
                      \t@countd=-1
                   \fi
	           \t@countc=\t@counta  \divide \t@countc by \t@countb
                   \t@counte=\t@countc  \multiply \t@counte by \t@countb
                   \advance \t@counta by -\t@counte
	           \t@counte=-1
                   \loop
                     \advance \t@counte by 1
	             \ifnum \t@counte<16
                       \multiply \t@countc by 2           
                       \multiply \t@counta by 2           
                       \ifnum \t@counta<\t@countb \else   
                         \advance \t@countc by 1          
                         \advance \t@counta by -\t@countb 
                       \fi
                   \repeat
	           \divide \t@countb by 2         
	           \ifnum \t@counta<\t@countb     
                     \advance \t@countc by 1
                   \fi
                   \ifnum \t@countd<0             
                     \t@countc=-\t@countc
                   \fi
                   \dimen0=\t@countc sp           
                   \edef #3{\expandafter\c@lean\the\dimen0}}

\outer\def\gnewif #1{\count@\escapechar \escapechar\m@ne
  \expandafter\expandafter\expandafter
   \edef\@if #1{true}{\global\let\noexpand#1=\noexpand\iftrue}%
  \expandafter\expandafter\expandafter
   \edef\@if #1{false}{\global\let\noexpand#1=\noexpand\iffalse}%
  \@if#1{false}\escapechar\count@} 
\def\@if #1#2{\csname\expandafter\if@\string#1#2\endcsname}
{\uccode`1=`i \uccode`2=`f \uppercase{\gdef\if@12{}}} 


\def\coordtopix #1#2{\dimen0=#1\d@dim
                     \dimen2=\d@sc\dimen0
                     \t@counta=\dimen2              
                     \t@countb=\s@ppix
                     \divide \t@countb by 2
                     \ifnum \t@counta<0             
                       \advance \t@counta by -\t@countb
                     \else
                       \advance \t@counta by \t@countb
                     \fi
                     \divide \t@counta by \s@ppix
                     #2=\t@counta}

\def\pixtocoord #1#2{\t@counta=#1%
                     \multiply \t@counta by \s@ppix
                     \dimen0=\d@sc\d@dim
                     \t@countb=\dimen0
                     \intdiv \t@counta \t@countb #2}

\def\pixtodim #1#2{\t@countb=#1%
                   \multiply \t@countb by \s@ppix
                   #2=\t@countb sp\relax}

\def\pixtobp #1#2{\dimen0=\p@sfactor pt
                  \t@counta=\dimen0
                  \multiply \t@counta by #1%
                  \ifnum \t@counta < 0             
                    \advance \t@counta by -32768
                  \else
                    \advance \t@counta by 32768
                  \fi
                  \divide \t@counta by 65536
                  #2=\t@counta}
                  
\newcount\t@counta    \newcount\t@countb   
\newcount\t@countc    \newcount\t@countd
\newcount\t@counte
\newcount\t@pixa      \newcount\t@pixb     
\newcount\t@pixc      \newcount\t@pixd
\let\l@lxbp=\t@pixa   \let\l@lybp=\t@pixb  
\let\u@rxbp=\t@pixc   \let\u@rybp=\t@pixd

\newdimen\t@xpos      \newdimen\t@ypos
\let\l@lxpos=\t@xpos  \let\l@lypos=\t@ypos

\newcount\xminpix      \newcount\xmaxpix
\newcount\yminpix      \newcount\ymaxpix

\newcount\a@lenpix     \newcount\a@widpix

\newcount\x@pix        \newcount\y@pix
\newcount\x@segoffpix  \newcount\y@segoffpix
\newcount\x@savepix    \newcount\y@savepix

\newcount\s@ppix       

\newcount\d@bs

\newcount\t@xdnum
\global\t@xdnum=0

\newdimen\hdrawsize    \newdimen\vdrawsize

\newbox\t@xdbox

\newwrite\drawfile

\newif\ifm@pending
\newif\ifp@ath
\newif\ifa@st
\gnewif \ift@extonly
\gnewif\ifp@osinit

\def\l@paren{(}
\def\a@st{*}

\catcode`\%=12
  \def\p@b {
\catcode`\%=14
\catcode`\{=12  \catcode`\}=12  \catcode`\u=1 \catcode`\v=2
  \def\l@br u{v  \def\r@br u}v
\catcode `\{=1  \catcode`\}=2   \catcode`\u=11 \catcode`\v=11

{\catcode`\p=12 \catcode`\t=12
 \gdef\c@lean #1pt{#1}}

\def\sppix#1/#2 {\dimen0=1#2 \s@ppix=\dimen0
                 \t@counta=#1%
                 \divide \t@counta by 2
                 \advance \s@ppix by \t@counta
                 \divide \s@ppix by #1
                 \t@counta=\s@ppix
                 \multiply \t@counta by 65536       
                 \advance \t@counta by 32891        
                 \divide \t@counta by 65782         
                 \dimen0=\t@counta sp
                 \edef\p@sfactor {\expandafter\c@lean\the\dimen0}}

\def\g@etargxy #1 #2 #3 #4\\#5#6{\def #5{#1}%
                                 \ifx #5\empty
                                   \g@etargxy #2 #3 #4 \\#5#6
                                 \else
                                   \def #6{#2}%
                                   \def\next {#3}%
                                   \ifx \next\empty \else
                                     \errmessage {TeXdraw: invalid coordinate}%
                                   \fi
                                 \fi}

\def\c@heckast #1{\expandafter
                  \c@heckastll #1\\}
\def\c@heckastll #1#2\\{\def\testit {#1}%
                        \ifx \testit\a@st
                          \a@sttrue
                        \else
                          \a@stfalse
                        \fi}

\def\g@etsympix #1#2{\expandafter
                     \ifx \csname #1\endcsname \relax
                       \errmessage {TeXdraw: undefined symbolic coordinate}%
                     \fi
                     #2=\csname #1\endcsname}

\def\s@etcsn #1#2{\expandafter
                  \xdef\csname#1\endcsname {#2}}

\def\g@etitem #1 #2\\#3#4{\edef #4{#2}\edef #3{#1}}
\def\a@pppix #1#2{\edef\next {#1}%
                  \ifx \next\empty \else
                    \coordtopix {#1}\t@pixa
                    \ifx #2\empty
                      \edef #2{\the\t@pixa}%
                    \else
                      \edef #2{#2 \the\t@pixa}%
                    \fi
                  \fi}

\def\s@etpospix #1#2{\coordtopix {#1}\x@pix
                     \advance \x@pix by \x@segoffpix
                     \coordtopix {#2}\y@pix
                     \advance \y@pix by \y@segoffpix
                     \u@pdateminmax \x@pix \y@pix}

\def\r@elpospix #1#2{\coordtopix {#1}\t@pixa
                     \advance \x@pix by \t@pixa
                     \coordtopix {#2}\t@pixa
                     \advance \y@pix by \t@pixa
                     \u@pdateminmax \x@pix \y@pix}

\def\r@elupd #1#2{\t@counta=\x@pix
                  \advance\t@counta by #1%
                  \t@countb=\y@pix
                  \advance\t@countb by #2%
                  \u@pdateminmax \t@counta \t@countb}

\def\u@pdateminmax #1#2{\ifnum #1>\xmaxpix
                          \global\xmaxpix=#1%
                        \fi
                        \ifnum #1<\xminpix
                          \global\xminpix=#1%
                        \fi
                        \ifnum #2>\ymaxpix
                          \global\ymaxpix=#2%
                        \fi
                        \ifnum #2<\yminpix
                          \global\yminpix=#2%
                        \fi}

\def\maxhvpos {\t@pixa=\xmaxpix
               \advance \t@pixa by -\xminpix
               \pixtodim  \t@pixa {\dimen2}%
               \global\hdrawsize=\dimen2
               \t@pixa=\ymaxpix
               \advance \t@pixa by -\yminpix
               \pixtodim \t@pixa {\dimen2}%
               \global\vdrawsize=\dimen2\relax}

\def\savemove #1#2{\x@savepix=#1\y@savepix=#2%
                   \m@pendingtrue
                   \ifp@osinit \else
                     \p@osinittrue
                     \global\xminpix=\x@savepix \global\yminpix=\y@savepix
                     \global\xmaxpix=\x@savepix \global\ymaxpix=\y@savepix
                   \fi}

\def\flushmove {\p@osinittrue
                \ifm@pending
                  \writetx {\the\x@savepix\space \the\y@savepix\space mv}%
                  \m@pendingfalse
                  \p@athfalse
                \fi}

\def\flushbs {\loop
                \ifnum \d@bs>0
                  \writetx {bs}%
                  \global\advance \d@bs by -1
              \repeat}
               
\def\h@move #1#2 #3)#4{\move (#2 #3)%
                       \h@text {#4}}
\def\h@text #1{\pixtodim \x@pix \t@xpos
               \pixtodim \y@pix \t@ypos
               \vbox to 0pt{\normalbaselines
                            \t@stuff
                            \kern -\t@ypos
                            \hbox to 0pt{\l@stuff
                                         \kern \t@xpos
                                         \hbox {#1}%
                                         \kern -\t@xpos
                                         \r@stuff}%
                            \kern \t@ypos
                            \b@stuff\relax}}

\def\r@move td:#1 #2#3 #4)#5{\move (#3 #4)%
                             \r@text td:#1 {#5}}
\def\r@text td:#1 #2{\pixtodim \x@pix \t@xpos
                     \pixtodim \y@pix \t@ypos
                     \vbox to 0pt{\kern -\t@ypos
                                  \hbox to 0pt{\kern \t@xpos
                                               \rottxt{#1}{#2}%
                                               \hss}%
                                  \vss}}

\def\rottxt #1#2{\rotsclTeX{#1}{1}{1}{\z@sb{#2}}}%
\def\z@sb #1{\vbox to 0pt{\normalbaselines
                          \t@stuff
                          \hbox to 0pt{\l@stuff
                                       \hbox {#1}%
                                       \r@stuff}%
                          \b@stuff}}

\def\t@exdrawdef {\sppix 300/in            
                  \drawdim in              
                  \edef\u@nitsc {1}
                  \setsegscale 1           
                  \arrowheadsize l:0.16 w:0.08
                  \arrowheadtype t:T
                  \textref h:L v:B }


\def\writeps #1{\flushbs
                \flushmove
                \p@athtrue
                \writetx {#1}}
\def\writetx #1{\ift@extonly
                  \t@extonlyfalse
                  \t@dropen
                \fi
                \w@rps {#1}}
\def\w@rps #1{\immediate\write\drawfile {#1}}

\def\t@dropen {%
  \global\advance \t@xdnum by 1
  \ifnum \t@xdnum<10
    \xdef\p@sfile {\jobname.ps\the\t@xdnum}
  \else
    \xdef\p@sfile {\jobname.p\the\t@xdnum}
  \fi
  \immediate\openout\drawfile=\p@sfile
  \w@rps {\p@b PS-Adobe-3.0 EPSF-3.0}%
  \w@rps {\p@p BoundingBox: (atend)}%
  \w@rps {\p@p Title: TeXdraw drawing: \p@sfile}%
  \w@rps {\p@p Pages: 1 1}%
  \w@rps {\p@p Creator: TeXdraw V1R3}%
  \w@rps {\p@p CreationDate: \the\year/\the\month/\the\day}%
  \w@rps {\p@p DocumentSuppliedResources: ProcSet TeXDraw 2.2 2}%
  \w@rps {\p@p DocumentData: Clean7Bit}%
  \w@rps {\p@p EndComments}%
  \w@rps {\p@p BeginDefaults}%
  \w@rps {\p@p PageNeededResources: ProcSet TeXDraw 2.2 2}%
  \w@rps {\p@p EndDefaults}%
  \w@rps {\p@p BeginProlog}%
  \w@rps {\p@p BeginResource: ProcSet TeXDraw 2.2 2 14696 10668}%
  \w@rps {\p@p VMlocation: local}%
  \w@rps {\p@p VMusage: 14696 10668}%
  \w@rps { /product where}%
  \w@rps {  {pop product (ghostscript) eq /setglobal {pop} def} if}%
  \w@rps { /setglobal where}%
  \w@rps {  {pop currentglobal false setglobal} if}%
  \w@rps { /setpacking where}%
  \w@rps {  {pop currentpacking false setpacking} if}%
  \w@rps {29 dict dup begin}%
  \w@rps {62 dict dup begin}%
  \w@rps { /rad 0 def /radx 0 def /rady 0 def /svm matrix def}%
  \w@rps { /hhwid 0 def /hlen 0 def /ah 0 def /tipy 0 def}%
  \w@rps { /tipx 0 def /taily 0 def /tailx 0 def /dx 0 def}%
  \w@rps { /dy 0 def /alen 0 def /blen 0 def}%
  \w@rps { /i 0 def /y1 0 def /x1 0 def /y0 0 def /x0 0 def}%
  \w@rps { /movetoNeeded 0 def}%
  \w@rps { /y3 0 def /x3 0 def /y2 0 def /x2 0 def}%
  \w@rps { /p1y 0 def /p1x 0 def /p2y 0 def /p2x 0 def}%
  \w@rps { /p0y 0 def /p0x 0 def /p3y 0 def /p3x 0 def}%
  \w@rps { /n 0 def /y 0 def /x 0 def}%
  \w@rps { /anglefactor 0 def /elemlength 0 def /excursion 0 def}%
  \w@rps { /endy 0 def /endx 0 def /beginy 0 def /beginx 0 def}%
  \w@rps { /centery 0 def /centerx 0 def /startangle 0 def }%
  \w@rps { /startradius 0 def /endradius 0 def /elemcount 0 def}%
  \w@rps { /smallincrement 0 def /angleincrement 0 def /radiusincrement 0 def}%
  \w@rps { /ifleft false def /ifright false def /iffill false def}%
  \w@rps { /freq 1 def /angle 0 def /yrad 0 def /xrad 0 def /y 0 def /x 0 def}%
  \w@rps { /saved 0 def}%
  \w@rps {end}%
  \w@rps {/dbdef {1 index exch 0 put 0 begin bind end def}}%
  \w@rps {  dup 3 4 index put dup 5 4 index put bind def pop}%
  \w@rps {/bdef {bind def} bind def}%
  \w@rps {/mv {stroke moveto} bdef}%
  \w@rps {/lv {lineto} bdef}%
  \w@rps {/st {currentpoint stroke moveto} bdef}%
  \w@rps {/sl {st setlinewidth} bdef}%
  \w@rps {/sd {st 0 setdash} bdef}%
  \w@rps {/sg {st setgray} bdef}%
  \w@rps {/bs {gsave} bdef /es {stroke grestore} bdef}%
  \w@rps {/cv {curveto} bdef}%
  \w@rps {/cr \l@br 0 begin}%
  \w@rps { gsave /rad exch def currentpoint newpath rad 0 360 arc}%
  \w@rps { stroke grestore end\r@br\space 0 dbdef}%
  \w@rps {/fc \l@br 0 begin}%
  \w@rps { gsave /rad exch def setgray currentpoint newpath}%
  \w@rps { rad 0 360 arc fill grestore end\r@br\space 0 dbdef}%
  \w@rps {/ar {gsave currentpoint newpath 5 2 roll arc stroke grestore} bdef}%
  \w@rps {/el \l@br 0 begin gsave /rady exch def /radx exch def}%
  \w@rps { svm currentmatrix currentpoint translate}%
  \w@rps { radx rady scale newpath 0 0 1 0 360 arc}%
  \w@rps { setmatrix stroke grestore end\r@br\space 0 dbdef}%
  \w@rps {/fl \l@br gsave closepath setgray fill grestore}%
  \w@rps { currentpoint newpath moveto\r@br\space bdef}%
  \w@rps {/fp \l@br gsave closepath setgray fill grestore}%
  \w@rps { currentpoint stroke moveto\r@br\space bdef}%
  \w@rps {/av \l@br 0 begin /hhwid exch 2 div def /hlen exch def}%
  \w@rps { /ah exch def /tipy exch def /tipx exch def}%
  \w@rps { currentpoint /taily exch def /tailx exch def}%
  \w@rps { /dx tipx tailx sub def /dy tipy taily sub def}%
  \w@rps { /alen dx dx mul dy dy mul add sqrt def}%
  \w@rps { /blen alen hlen sub def}%
  \w@rps { gsave tailx taily translate dy dx atan rotate}%
  \w@rps { (V) ah ne {blen 0 gt {blen 0 lineto} if} {alen 0 lineto} ifelse}%
  \w@rps { stroke blen hhwid neg moveto alen 0 lineto blen hhwid lineto}%
  \w@rps { (T) ah eq {closepath} if}%
  \w@rps { (W) ah eq {gsave 1 setgray fill grestore closepath} if}%
  \w@rps { (F) ah eq {fill} {stroke} ifelse}%
  \w@rps { grestore tipx tipy moveto end\r@br\space 0 dbdef}%
  \w@rps {/setupcurvy \l@br 0 begin}%
  \w@rps { dup 0 eq {1 add} if /anglefactor exch def}%
  \w@rps { abs dup 0 eq {1 add} if /elemlength exch def /excursion exch def}%
  \w@rps { /endy exch def /endx exch def}%
  \w@rps { /beginy exch def /beginx exch def}%
  \w@rps { /centery exch def /centerx exch def}%
  \w@rps { cleartomark}%
  \w@rps { /startangle beginy centery sub beginx centerx sub atan def}%
  \w@rps { /startradius beginy centery sub dup mul }%
  \w@rps {              beginx centerx sub dup mul add sqrt def}%
  \w@rps { /endradius endy centery sub dup mul }%
  \w@rps {            endx centerx sub dup mul add sqrt def}%
  \w@rps { endradius startradius sub }%
  \w@rps { endy centery sub endx centerx sub atan }%
  \w@rps { startangle 2 copy le {exch 360 add exch} if sub dup}%
  \w@rps { elemlength startradius endradius add atan dup add}%
  \w@rps { div round abs cvi dup 0 eq {1 add} if}%
  \w@rps { dup /elemcount exch def }%
  \w@rps { div dup anglefactor div dup /smallincrement exch def}%
  \w@rps { sub /angleincrement exch def}%
  \w@rps { elemcount div /radiusincrement exch def}%
  \w@rps { gsave newpath}%
  \w@rps { startangle dup cos startradius mul }%
  \w@rps { centerx add exch }%
  \w@rps { sin startradius mul centery add moveto}%
  \w@rps { end \r@br 0 dbdef}%
  \w@rps {/curvyphoton \l@br 0 begin}%
  \w@rps { setupcurvy}%
  \w@rps { elemcount \l@br /startangle startangle smallincrement add def}%
  \w@rps {            /startradius startradius excursion add def}%
  \w@rps {            startangle dup cos startradius mul }%
  \w@rps {            centerx add exch }%
  \w@rps {            sin startradius mul centery add}%
  \w@rps {	      /excursion excursion neg def}%
  \w@rps {	      /startangle startangle angleincrement add }%
  \w@rps {                        smallincrement sub def}%
  \w@rps {	      /startradius startradius radiusincrement add def}%
  \w@rps {	      startangle dup cos startradius mul }%
  \w@rps {	      centerx add exch }%
  \w@rps {            sin startradius mul centery add}%
  \w@rps {	      /startradius startradius excursion add def}%
  \w@rps {            /startangle startangle smallincrement add def}%
  \w@rps {             startangle dup cos startradius mul }%
  \w@rps {	       centerx add exch }%
  \w@rps {             sin startradius mul centery add curveto\r@br repeat}%
  \w@rps {	       stroke grestore end}%
  \w@rps {	      \r@br 0 dbdef}%
  \w@rps {/curvygluon \l@br 0 begin}%
  \w@rps { setupcurvy /radiusincrement radiusincrement 2 div def}%
  \w@rps { elemcount \l@br startangle angleincrement add dup}%
  \w@rps {            cos startradius mul centerx add exch}%
  \w@rps {            sin startradius mul centery add}%
  \w@rps {            /startradius startradius radiusincrement add}%
  \w@rps {                         excursion sub def}%
  \w@rps {            startangle angleincrement add dup}%
  \w@rps {            cos startradius mul centerx add exch}%
  \w@rps {            sin startradius mul centery add}%
\w@rps{            startangle angleincrement smallincrement add 2 div add dup}%
  \w@rps {            cos startradius mul centerx add exch}%
  \w@rps {            sin startradius mul centery add}%
  \w@rps {	    curveto}%
\w@rps{      /startangle startangle angleincrement smallincrement add add def}%
  \w@rps {            startangle angleincrement sub dup}%
  \w@rps {            cos startradius mul centerx add exch}%
  \w@rps {            sin startradius mul centery add}%
  \w@rps {	    /startradius startradius radiusincrement add}%
  \w@rps {			excursion add def}%
  \w@rps {            startangle angleincrement sub dup}%
  \w@rps {            cos startradius mul centerx add exch}%
  \w@rps {            sin startradius mul centery add}%
  \w@rps {            startangle dup}%
  \w@rps {            cos startradius mul centerx add exch}%
  \w@rps {            sin startradius mul centery add}%
  \w@rps {	    curveto\r@br repeat}%
  \w@rps { stroke grestore end}%
  \w@rps { \r@br 0 dbdef}%
  \w@rps {/blob \l@br}%
  \w@rps {0 begin st gsave}%
  \w@rps {dup type dup}%
  \w@rps {/stringtype eq}%
  \w@rps {\l@br pop 0 get }%
  \w@rps {dup (B) 0 get eq dup 2 index}%
  \w@rps {(L) 0 get eq or /ifleft exch def}%
  \w@rps {exch (R) 0 get eq or /ifright exch def}%
  \w@rps {/iffill false def \r@br}%
  \w@rps {\l@br /ifleft false def}%
  \w@rps {/ifright false def}%
  \w@rps {/booleantype eq }%
  \w@rps {{/iffill exch def}}%
  \w@rps {{setgray /iffill true def} ifelse \r@br}%
  \w@rps {ifelse}%
  \w@rps {/freq exch def}%
  \w@rps {/angle exch def}%
  \w@rps {/yrad  exch def}%
  \w@rps {/xrad  exch def}%
  \w@rps {/y exch def}%
  \w@rps {/x exch def}%
  \w@rps {newpath}%
  \w@rps {svm currentmatrix pop}%
  \w@rps {x y translate 	}%
  \w@rps {angle rotate}%
  \w@rps {xrad yrad scale}%
  \w@rps {0 0 1 0 360 arc}%
  \w@rps {gsave 1 setgray fill grestore}%
  \w@rps {gsave svm setmatrix stroke grestore}%
  \w@rps {gsave iffill {fill} if grestore}%
  \w@rps {clip newpath}%
  \w@rps {gsave }%
  \w@rps {ifleft  \l@br -3 freq 3 { -1 moveto 2 2 rlineto} for}%
  \w@rps {svm setmatrix stroke\r@br if }%
  \w@rps {grestore}%
  \w@rps {ifright \l@br 3 freq neg -3 { -1 moveto -2 2 rlineto} for}%
  \w@rps {svm setmatrix stroke\r@br if}%
  \w@rps {grestore end}%
  \w@rps {\r@br 0 dbdef}%
  \w@rps {/BSpl \l@br}%
  \w@rps { 0 begin}%
  \w@rps { storexyn}%
  \w@rps { currentpoint newpath moveto}%
  \w@rps { n 1 gt \l@br}%
  \w@rps {  0 0 0 0 0 0 1 1 true subspline}%
  \w@rps {  n 2 gt \l@br}%
  \w@rps {   0 0 0 0 1 1 2 2 false subspline}%
  \w@rps {   1 1 n 3 sub \l@br}%
  \w@rps {    /i exch def}%
  \w@rps {    i 1 sub dup i dup i 1 add dup i 2 add dup false subspline}%
  \w@rps {    \r@br for}%
  \w@rps {   n 3 sub dup n 2 sub dup n 1 sub dup 2 copy false subspline}%
  \w@rps {   \r@br if}%
  \w@rps {  n 2 sub dup n 1 sub dup 2 copy 2 copy false subspline}%
  \w@rps {  \r@br if}%
  \w@rps { end}%
  \w@rps { \r@br 0 dbdef}%
  \w@rps {/midpoint \l@br}%
  \w@rps { 0 begin}%
  \w@rps { /y1 exch def}%
  \w@rps { /x1 exch def}%
  \w@rps { /y0 exch def}%
  \w@rps { /x0 exch def}%
  \w@rps { x0 x1 add 2 div}%
  \w@rps { y0 y1 add 2 div}%
  \w@rps { end}%
  \w@rps { \r@br 0 dbdef}%
  \w@rps {/thirdpoint \l@br}%
  \w@rps { 0 begin}%
  \w@rps { /y1 exch def}%
  \w@rps { /x1 exch def}%
  \w@rps { /y0 exch def}%
  \w@rps { /x0 exch def}%
  \w@rps { x0 2 mul x1 add 3 div}%
  \w@rps { y0 2 mul y1 add 3 div}%
  \w@rps { end}%
  \w@rps { \r@br 0 dbdef}%
  \w@rps {/subspline \l@br}%
  \w@rps { 0 begin}%
  \w@rps { /movetoNeeded exch def}%
  \w@rps { y exch get /y3 exch def}%
  \w@rps { x exch get /x3 exch def}%
  \w@rps { y exch get /y2 exch def}%
  \w@rps { x exch get /x2 exch def}%
  \w@rps { y exch get /y1 exch def}%
  \w@rps { x exch get /x1 exch def}%
  \w@rps { y exch get /y0 exch def}%
  \w@rps { x exch get /x0 exch def}%
  \w@rps { x1 y1 x2 y2 thirdpoint}%
  \w@rps { /p1y exch def}%
  \w@rps { /p1x exch def}%
  \w@rps { x2 y2 x1 y1 thirdpoint}%
  \w@rps { /p2y exch def}%
  \w@rps { /p2x exch def}%
  \w@rps { x1 y1 x0 y0 thirdpoint}%
  \w@rps { p1x p1y midpoint}%
  \w@rps { /p0y exch def}%
  \w@rps { /p0x exch def}%
  \w@rps { x2 y2 x3 y3 thirdpoint}%
  \w@rps { p2x p2y midpoint}%
  \w@rps { /p3y exch def}%
  \w@rps { /p3x exch def}%
  \w@rps { movetoNeeded \l@br p0x p0y moveto \r@br if}%
  \w@rps { p1x p1y p2x p2y p3x p3y curveto}%
  \w@rps { end}%
  \w@rps { \r@br 0 dbdef}%
  \w@rps {/storexyn \l@br}%
  \w@rps { 0 begin}%
  \w@rps { /n exch def}%
  \w@rps { /y n array def}%
  \w@rps { /x n array def}%
  \w@rps { n 1 sub -1 0 \l@br}%
  \w@rps {  /i exch def}%
  \w@rps {  y i 3 2 roll put}%
  \w@rps {  x i 3 2 roll put}%
  \w@rps {  \r@br for end}%
  \w@rps { \r@br 0 dbdef}%
  \w@rps {/bop \l@br save 0 begin /saved exch def end}%
  \w@rps { scale setlinecap setlinejoin setlinewidth setdash moveto}%
  \w@rps { \r@br 1 dbdef}%
  \w@rps {/eop {stroke 0 /saved get restore showpage} 1 dbdef}%
  \w@rps {end /defineresource where}%
  \w@rps { {pop mark exch /TeXDraw exch /ProcSet defineresource cleartomark}}%
  \w@rps { {/TeXDraw exch readonly def} ifelse}%
  \w@rps {/setpacking where {pop setpacking} if}%
  \w@rps {/setglobal where {pop setglobal} if}%
  \w@rps {\p@p EndResource}%
  \w@rps {\p@p EndProlog}%
  \w@rps {\p@p Page: 1 1}%
  \w@rps {\p@p PageBoundingBox: (atend)}%
  \w@rps {\p@p BeginPageSetup}%
  \w@rps {/TeXDraw /findresource where}%
  \w@rps { {pop /ProcSet findresource}}%
  \w@rps { {load} ifelse}%
  \w@rps {begin}%
  \w@rps {0 0 [] 0 3 1 1 \p@sfactor\space \p@sfactor\space bop}%
  \w@rps {\p@p EndPageSetup}%
}

\def\t@drclose {%
  \pixtobp \xminpix \l@lxbp  \pixtobp \yminpix \l@lybp
  \pixtobp \xmaxpix \u@rxbp  \pixtobp \ymaxpix \u@rybp
  \w@rps {\p@p PageTrailer}%
  \w@rps {\p@p PageBoundingBox: \the\l@lxbp\space \the\l@lybp\space
                            \the\u@rxbp\space \the\u@rybp}%
  \w@rps {eop end}%
  \w@rps {\p@p Trailer}%
  \w@rps {\p@p BoundingBox: \the\l@lxbp\space \the\l@lybp\space
                            \the\u@rxbp\space \the\u@rybp}%
  \w@rps {\p@p EOF}%
  \closeout\drawfile
}

\catcode`\@=\catamp
\def\dvialwsetup{
\def\includepsfile##1##2##3##4##5{\special{Insert ##1\space%
                                 }}%
\def\rotsclTeX##1##2##3##4{\special{Insert /dev/null do %
                              3 index exch translate cleartomark %
                              matrix currentmatrix aload pop %
                              7 6 roll restore matrix astore %
                              matrix currentmatrix exch setmatrix %
                              0 0 moveto setmatrix %
                              gsave currentpoint 2 copy translate ##1 rotate %
                              ##2 ##3 scale neg exch neg exch translate %
                              save}%
                   ##4%
                   \special{Insert /dev/null do cleartomark restore %
                           {currentpoint} stopped grestore {moveto} save}}%
}
\def\dvipssetup{
\def\includepsfile##1##2##3##4##5{\vbox to 0pt{%
                             \vskip##5%
			     \includegraphics{##1}%
                             \vss}}
\def\rotsclTeX##1##2##3##4{%
		       ##4
}
}
\dvipssetup

\expandafter\ifx\csname fonts are loaded\endcsname\relax\else \fi
\immediate\openin0 localfonts.tex
\ifeof0\relax \else\ifx\localfontsloaded\donotdefinethis\else \fi
\font\seventeenrm=cmr17 \font\twelverm=cmr12 \font\tenrm=cmr10 
\font\ninerm=cmr9       \font\eightrm=cmr8   \font\sevenrm=cmr7
\font\sixrm=cmr6        \font\fiverm=cmr5
\font\twentyfourrm=cmr17 at 24pt \font\twentyrm=cmr17 at 20pt
\font\sixteenrm=cmr17 at 16pt \font\fourteenrm=cmr12 at 14pt
\font\twelvei=cmmi12    \font\teni=cmmi10    \font\ninei=cmmi9
\font\eighti=cmmi8      \font\seveni=cmmi7   \font\sixi=cmmi6
\font\fivei=cmmi5
\font\twentyfouri=cmmi12 at 24pt \font\twentyi=cmmi12 at 20pt
\font\sixteeni=cmmi12 at 16pt \font\fourteeni=cmmi12 at 14pt
\font\tensy=cmsy10      \font\ninesy=cmsy9   \font\eightsy=cmsy8
\font\sevensy=cmsy7     \font\sixsy=cmsy6    \font\fivesy=cmsy5
\skewchar\tensy='60 \skewchar\ninesy='60 \skewchar\eightsy='60
\skewchar\sevensy='60 \skewchar\sixsy='60 \skewchar\fivesy='60
\font\tenex=cmex10      \font\nineex=cmex9   \font\eightex=cmex8
\font\sevenex=cmex7
\font\fiveex=cmex7 at 5pt
\font\twelveit=cmti12   \font\tenit=cmti10   \font\nineit=cmti9
\font\eightit=cmti8     \font\sevenit=cmti7
\font\twentyfourit=cmti12 at 24pt \font\twentyit=cmti12 at 20pt
\font\sixteenit=cmti12 at 16pt \font\fourteenit=cmti12 at 14pt
\font\fiveit=cmti7 at 5pt
\font\twelvesl=cmsl12   \font\tensl=cmsl10   \font\ninesl=cmsl9
\font\eightsl=cmsl8
\font\twentyfoursl=cmsl12 at 24pt \font\twentysl=cmsl12 at 20pt
\font\sixteensl=cmsl12 at 16pt \font\fourteensl=cmsl12 at 16pt
\font\sevensl=cmsl8 at 7pt \font\fivesl=cmsl8 at 5pt
\font\twelvett=cmtt12   \font\tentt=cmtt10   \font\ninett=cmtt9
\font\eighttt=cmtt8
\hyphenchar\twelvett=-1 \hyphenchar\tentt=-1 \hyphenchar\ninett=-1
\hyphenchar\eighttt=-1
\font\twentyfourtt=cmtt12 at 24pt \font\twentytt=cmtt12 at 20pt
\font\sixteentt=cmtt12 at 16pt \font\fourteentt=cmtt12 at 16pt
\font\seventt=cmtt8 at 7pt \font\fivett=cmtt8 at 5pt
\hyphenchar\twentyfourtt=-1 \hyphenchar\twentytt=-1
\hyphenchar\sixteentt=-1 \hyphenchar\fourteentt=-1
\hyphenchar\seventt=-1 \hyphenchar\fivett=-1
\font\tenbf=cmb10
\font\seventeenss=cmss17\font\twelvess=cmss12\font\tenss=cmss10
\font\niness=cmss9      \font\eightss=cmss8
\font\twentyfourss=cmss17 at 24pt \font\twentyss=cmss17 at 20pt
\font\sixteenss=cmss17 at 16pt \font\fourteenss=cmss12 at 14pt
\font\sevenss=cmss8 at 7pt \font\fivess=cmss8 at 5pt
\font\tenmib=cmmib10    \font\ninemib=cmmib9 \font\eightmib=cmmib8
\font\sevenmib=cmmib7   \font\sixmib=cmmib6  \font\fivemib=cmmib5
\skewchar\tenmib='177   \skewchar\ninemib='177 \skewchar\eightmib='177
\skewchar\sevenmib='177 \skewchar\sixmib='177  \skewchar\fivemib='177
\font\tenbsy=cmbsy10    \font\ninebsy=cmbsy9 \font\eightbsy=cmbsy8
\font\sevenbsy=cmbsy7   \font\sixbsy=cmbsy6  \font\fivebsy=cmbsy5
\skewchar\tenbsy='60 \skewchar\ninebsy='60 \skewchar\eightbsy='60
\skewchar\sevenbsy='60 \skewchar\sixbsy='60 \skewchar\fivebsy='60
\let\localfontsloaded=\relax
\fi
\newif\ifplain
\plainfalse
\def\loadfonts#1#2{%
 \expandafter\ifx\csname#1rm\endcsname\relax
 \expandafter\global\expandafter\font\csname#1rm\endcsname=cmr10 at#2pt \fi
 \expandafter\ifx\csname#1i\endcsname\relax
 \expandafter\global\expandafter\font\csname#1i\endcsname =cmmi10 at#2pt 
 \skewchar\csname#1i\endcsname ='177 \fi
 \expandafter\ifx\csname#1sy\endcsname\relax
 \expandafter\global\expandafter\font\csname#1sy\endcsname=cmsy10 at#2pt 
 \skewchar\csname#1sy\endcsname= '60 \fi
 \expandafter\ifx\csname#1ex\endcsname\relax
 \expandafter\global\expandafter\font\csname#1ex\endcsname=cmex10 at#2pt \fi
 \expandafter\ifx\csname#1it\endcsname\relax
 \expandafter\global\expandafter\font\csname#1it\endcsname=cmti10 at#2pt \fi
 \expandafter\ifx\csname#1sl\endcsname\relax
 \expandafter\global\expandafter\font\csname#1sl\endcsname=cmsl10 at#2pt \fi
 \expandafter\ifx\csname#1tt\endcsname\relax
 \expandafter\global\expandafter\font\csname#1tt\endcsname=cmtt10 at#2pt 
 \hyphenchar\csname#1tt\endcsname=  -1 \fi
 \expandafter\ifx\csname#1bf\endcsname\relax
 \expandafter\global\expandafter\font\csname#1bf\endcsname=cmb10  at#2pt \fi
 \expandafter\ifx\csname#1ss\endcsname\relax
 \expandafter\global\expandafter\font\csname#1ss\endcsname=cmss10 at#2pt \fi
 \expandafter\ifx\csname#1mib\endcsname\relax
 \expandafter\global\expandafter\font\csname#1mib\endcsname=cmmib10 at #2pt 
 \skewchar\csname#1mib\endcsname='177 \fi
 \expandafter\ifx\csname#1bsy\endcsname\relax
 \expandafter\global\expandafter\font\csname#1bsy\endcsname=cmbsy10 at #2pt 
 \skewchar\csname#1bsy\endcsname= '60 \fi
}

\ifplain 
\expandafter\ifx\csname tenmib\endcsname\relax\global\font\tenmib=cmmib10 \fi
\expandafter\ifx\csname ninemib\endcsname\global\font\ninemib=cmmib10 at9pt\fi
\expandafter\ifx\csname sevenmi\endcsname\global\font\sevenmib=cmmib10 at7pt\fi
\expandafter\ifx\csname fivemib\endcsname\global\font\fivemib=cmmib10 at5pt\fi
\expandafter\ifx\csname tenbsy\endcsname\global\font\tenbsy=cmbsy10 \fi
\expandafter\ifx\csname ninebsy\endcsname\global\font\ninebsy=cmbsy10 at9pt\fi
\expandafter\ifx\csname sevenbs\endcsname\global\font\sevenbsy=cmbsy10 at7pt\fi
\expandafter\ifx\csname fivebsy\endcsname\global\font\fivebsy=cmbsy10 at5pt\fi
\else
\def\tenfonts{%
	\loadfonts{ten}{10}%
	\global\def\tenfonts{}}
\def\ninefonts{%
	\loadfonts{nine}{9}%
	\global\def\ninefonts{}}
\def\sevenfonts{%
	\loadfonts{seven}{7}%
	\global\def\sevenfonts{}}
\def\fivefonts{%
	\loadfonts{five}{5}%
	\global\def\fivefonts{}}
\fi
\def\twelvefonts{%
	\loadfonts{twelve}{12}%
	\global\def\twelvefonts{}}
\def\fourteenfonts{%
	\loadfonts{fourteen}{14}%
	\global\def\fourteenfonts{}}
\def\sixteenfonts{%
	\loadfonts{sixteen}{16}%
	\global\def\sixteenfonts{}}
\def\twentyfonts{%
	\loadfonts{twenty}{20}%
	\global\def\twentyfonts{}}
\def\twentyfourfonts{%
	\loadfonts{twentyfour}{24}%
	\global\def\twentyfourfonts{}}
\def\famset#1#2#3#4#5{%
	\textfont#1\csname#3#2\endcsname
	\scriptfont#1\csname#4#2\endcsname
	\scriptscriptfont#1\csname#5#2\endcsname}
\def\ninepoint{%
	\ifplain\else\ninefonts\sevenfonts\fivefonts\fi
	\famset0{rm}{nine}{seven}{five}%
	\famset1{i} {nine}{seven}{five}%
	\famset2{sy}{nine}{seven}{five}%
	\famset3{ex}{nine}{seven}{five}%
	\famset\itfam{it}{nine}{seven}{five}%
	\famset\slfam{sl}{nine}{seven}{five}%
	\famset\ttfam{tt}{ten}{seven}{five}%
	\famset\bffam{bf}{nine}{seven}{five}%
	\def\rm{\fam0\ninerm}%
	\def\it{\fam\itfam\nineit}%
	\def\sl{\fam\slfam\ninesl}%
	\def\tt{\fam\ttfam\tentt}%
	\def\bf{\famset0{bf}{nine}{seven}{five}%
		\famset1{mib}{nine}{seven}{five}%
		\famset2{bsy}{nine}{seven}{five}%
		\fam\bffam\ninebf}%
	\setbox\strutbox=\hbox{\vrule height 8pt depth 3pt width 0pt}%
	\baselineskip11pt\rm%
	}
\def\tenpoint{%
	\ifplain\else\tenfonts\sevenfonts\fivefonts\fi
	\famset0{rm}{ten}{seven}{five}%
	\famset1{i} {ten}{seven}{five}%
	\famset2{sy}{ten}{seven}{five}%
	\famset3{ex}{ten}{seven}{five}%
	\famset\itfam{it}{ten}{seven}{five}%
	\famset\slfam{sl}{ten}{seven}{five}%
	\famset\ttfam{tt}{ten}{seven}{five}%
	\famset\bffam{bf}{ten}{seven}{five}%
	\def\rm{\fam0\tenrm}%
	\def\it{\fam\itfam\tenit}%
	\def\sl{\fam\slfam\tensl}%
	\def\tt{\fam\ttfam\tentt}%
	\def\bf{\famset0{bf}{ten}{seven}{five}%
		\famset1{mib}{ten}{seven}{five}%
		\famset2{bsy}{ten}{seven}{five}%
		\fam\bffam\tenbf}%
	\setbox\strutbox=\hbox{\vrule height 8.5pt depth 3.5pt width 0pt}%
	\baselineskip12pt\rm%
	}
\def\twelvepoint{%
	\twelvefonts\ifplain\else\ninefonts\sevenfonts\fi
	\famset0{rm}{twelve}{nine}{seven}%
	\famset1{i} {twelve}{nine}{seven}%
	\famset2{sy}{twelve}{nine}{seven}%
	\famset3{ex}{twelve}{nine}{seven}%
	\famset\itfam{it}{twelve}{nine}{seven}%
	\famset\slfam{sl}{twelve}{nine}{seven}%
	\famset\ttfam{tt}{twelve}{nine}{seven}%
	\famset\bffam{bf}{twelve}{nine}{seven}%
	\def\rm{\fam0\twelverm}%
	\def\it{\fam\itfam\twelveit}%
	\def\sl{\fam\slfam\twelvesl}%
	\def\tt{\fam\ttfam\twelvett}%
	\def\bf{\famset0{bf}{twelve}{nine}{seven}%
		\famset1{mib}{twelve}{nine}{seven}%
		\famset2{bsy}{twelve}{nine}{seven}%
		\fam\bffam\twelvebf}%
	\setbox\strutbox=\hbox{\vrule height 10pt depth 4pt width 0pt}%
	\baselineskip14pt\rm%
	}
\def\fourteenpoint{%
	\fourteenfonts\twelvefonts\ifplain\else\tenfonts\fi
	\famset0{rm}{fourteen}{twelve}{ten}%
	\famset1{i} {fourteen}{twelve}{ten}%
	\famset2{sy}{fourteen}{twelve}{ten}%
	\famset3{ex}{fourteen}{twelve}{ten}%
	\famset\itfam{it}{fourteen}{twelve}{ten}%
	\famset\slfam{sl}{fourteen}{twelve}{ten}%
	\famset\ttfam{tt}{fourteen}{twelve}{ten}%
	\famset\bffam{bf}{fourteen}{twelve}{ten}%
	\def\rm{\fam0\fourteenrm}%
	\def\it{\fam\itfam\fourteenit}%
	\def\sl{\fam\slfam\fourteensl}%
	\def\tt{\fam\ttfam\fourteentt}%
	\def\bf{\famset0{bf}{fourteen}{twelve}{ten}%
		\famset1{mib}{fourteen}{twelve}{ten}%
		\famset2{bsy}{fourteen}{twelve}{ten}%
		\fam\bffam\fourteenbf}%
	\setbox\strutbox=\hbox{\vrule height 12pt depth 5pt width 0pt}%
	\baselineskip17pt\rm%
	}
\def\sixteenpoint{%
	\sixteenfonts\fourteenfonts\twelvefonts
	\famset0{rm}{sixteen}{fourteen}{twelve}%
	\famset1{i} {sixteen}{fourteen}{twelve}%
	\famset2{sy}{sixteen}{fourteen}{twelve}%
	\famset3{ex}{sixteen}{fourteen}{twelve}%
	\famset\itfam{it}{sixteen}{fourteen}{twelve}%
	\famset\slfam{sl}{sixteen}{fourteen}{twelve}%
	\famset\ttfam{tt}{sixteen}{fourteen}{twelve}%
	\famset\bffam{bf}{sixteen}{fourteen}{twelve}%
	\def\rm{\fam0\sixteenrm}%
	\def\it{\fam\itfam\sixteenit}%
	\def\sl{\fam\slfam\sixteensl}%
	\def\tt{\fam\ttfam\sixteentt}%
	\def\bf{\famset0{bf}{sixteen}{fourteen}{twelve}%
		\famset1{mib}{sixteen}{fourteen}{twelve}%
		\famset2{bsy}{sixteen}{fourteen}{twelve}%
		\fam\bffam\sixteenbf}%
	\setbox\strutbox=\hbox{\vrule height 14pt depth 6pt width 0pt}%
	\baselineskip20pt\rm%
	}
\def\twentypoint{%
	\twentyfonts\sixteenfonts\fourteenfonts
	\famset0{rm}{twenty}{sixteen}{fourteen}%
	\famset1{i} {twenty}{sixteen}{fourteen}%
	\famset2{sy}{twenty}{sixteen}{fourteen}%
	\famset3{ex}{twenty}{sixteen}{fourteen}%
	\famset\itfam{it}{twenty}{sixteen}{fourteen}%
	\famset\slfam{sl}{twenty}{sixteen}{fourteen}%
	\famset\ttfam{tt}{twenty}{sixteen}{fourteen}%
	\famset\bffam{bf}{twenty}{sixteen}{fourteen}%
	\def\rm{\fam0\twentyrm}%
	\def\it{\fam\itfam\twentyit}%
	\def\sl{\fam\slfam\twentysl}%
	\def\tt{\fam\ttfam\twentytt}%
	\def\bf{\famset0{bf}{twenty}{sixteen}{fourteen}%
		\famset1{mib}{twenty}{sixteen}{fourteen}%
		\famset2{bsy}{twenty}{sixteen}{fourteen}%
		\fam\bffam\twentybf}%
	\setbox\strutbox=\hbox{\vrule height 17pt depth 7pt width 0pt}%
	\baselineskip24pt\rm%
	}
\def\twentyfourpoint{%
	\twentyfourfonts\twentyfonts\sixteenfonts
	\famset0{rm}{twentyfour}{twenty}{sixteen}%
	\famset1{i} {twentyfour}{twenty}{sixteen}%
	\famset2{sy}{twentyfour}{twenty}{sixteen}%
	\famset3{ex}{twentyfour}{twenty}{sixteen}%
	\famset\itfam{it}{twentyfour}{twenty}{sixteen}%
	\famset\slfam{sl}{twentyfour}{twenty}{sixteen}%
	\famset\ttfam{tt}{twentyfour}{twenty}{sixteen}%
	\famset\bffam{bf}{twentyfour}{twenty}{sixteen}%
	\def\rm{\fam0\twentyfourrm}%
	\def\it{\fam\itfam\twentyfourit}%
	\def\sl{\fam\slfam\twentyfoursl}%
	\def\tt{\fam\ttfam\twentyfourtt}%
	\def\bf{\famset0{bf}{twentyfour}{twenty}{sixteen}%
		\famset1{mib}{twentyfour}{twenty}{sixteen}%
		\famset2{bsy}{twentyfour}{twenty}{sixteen}%
		\fam\bffam\twentyfourbf}%
	\setbox\strutbox=\hbox{\vrule height 17pt depth 7pt width 0pt}%
	\baselineskip24pt\rm%
	}
\expandafter\def\csname fonts are loaded\endcsname{}%

\newif\ifintexdraw
\ifx\axisscale\donotdefinethis\def\axisscale{1}\fi
\def\e{\ifintexdraw\immediate\message{Ending figure}%
       \esegment\etexdraw\ifvmode\medskip\fi\intexdrawfalse\else
       \immediate\message{\string\e\space ignored}\fi}
\def\m#1#2{\move (#1 #2)}
\def\w#1{\linewd #1 }
\def\p#1#2{\move (#1 #2) \fcir f:0 r:1}
\def\l#1#2#3#4{\move (#1 #2) \lvec (#3 #4)}
\def\n#1#2{\lvec (#1 #2)}
\def\s#1#2#3#4{\ifintexdraw\immediate\message{\string\s\space ignored}%
               \else\intexdrawtrue
	       \immediate\message{Starting figure}
               \btexdraw
               \drawdim bp %
               \setunitscale 0.08791 %
	       \bsegment
	       \expandafter\expandafter\expandafter
               \relsegscale\expandafter\axisscale\space
               \fi}
\def\TX{\ifmmode\else$\fi\rm}
\def\XT{\ifmmode$\relax\else\fi}
{\catcode`p=12 \catcode`t=12
 \gdef\dimno#1pt{#1}}%
\def\scle#1#2{{{\dimen0=1000pt
                \dimen0=#2\dimen0\relax
                \expandafter\dimen\expandafter0\expandafter=
                \axisscale\dimen0\relax\divide\dimen0 by 1000\relax
                \xdef\myowntemp{\expandafter\dimno\the\dimen0}}
               \setbox0\hbox{\rotsclTeX{0}{\myowntemp}{\myowntemp}{\hbox{#1}}}%
               \ht0=\myowntemp\ht0
	       \dp0=\myowntemp\dp0
	       \wd0=\myowntemp\wd0
               \box0}}
\def\t#1{\textref h:L v:B \htext{$\rm #1$}}
\def\ltx(#1 #2) #3#4{\textref h:L v:C \htext (#1 #2){\scle{#3\XT}{#4}}}
\def\rtx(#1 #2) #3#4{\textref h:R v:C \htext (#1 #2){\scle{#3\XT}{#4}}}
\def\ctx(#1 #2) #3#4{\textref h:C v:C \htext (#1 #2){\scle{#3\XT}{#4}}}
\def\vtx(#1 #2) #3#4{\textref h:C v:T \vtext (#1 #2){\scle{#3\XT}{#4}}}
\def\solid{\lpatt ()}
\def\disconnected{\lpatt (0 100)}
\def\dotted{\lpatt (10 30)}
\def\dotdashed{\lpatt (10 40 100 100)}
\def\shortdashed{\lpatt (100 100)}
\def\longdashed{\lpatt (200 100)}
\def\dashdotdotted{\lpatt (100 100 10 40 10 40)}
\def\f#1{%
\csname#1\endcsname
}
\let\axisloaded=\relax

\begin{figure}[t]
\def\axisscale{0.7}
\input {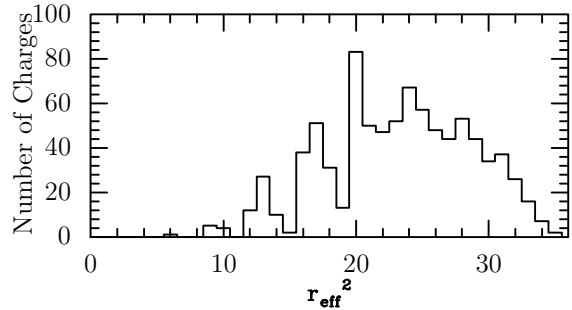}
\vspace{-30pt}
\caption{Total Number of charges for IA with $q_{min}(\tau=10)=1$
versus the effective radius $r_{eff}^2$.}
\label{fig:reffch}
\end{figure}

\begin{table}[b]
\caption{Improved action $\chi_t^{1/4}$ (MeV) (28 lattices)}
\setlength{\tabcolsep}{6pt}
\begin{tabular}{|r|c|c|l|l|}
\hline
\multicolumn{1}{|c|}{} &
\multicolumn{1}{|c|}{} &
\multicolumn{1}{|c|}{Geometric} &
\multicolumn{2}{|c|}{Plaquette} \cr 
\multicolumn{1}{|c|}{$\tau$} & 
\multicolumn{1}{|c|}{$\langle \Box \rangle$} & 
\multicolumn{1}{|c|}{$\chi_t^{1/4}$} &
\multicolumn{1}{|c|}{$Z_Q^{1/2}\chi_t^{1/4}$} &
\multicolumn{1}{|c|}{$\chi_t^{1/4}$} \cr \hline
$ 0$ & $1.871$ & $238(14)$  &  $170(9) $ & $354(19)$ \cr
$ 1$ & $2.687$ & $238(14)$  &  $172(10)$ & $236(14)$ \cr
$ 3$ & $2.922$ & $238(15)$  &  $196(13)$ & $229(15)$ \cr
$ 6$ & $2.970$ & $238(16)$  &  $211(13)$ & $241(15)$ \cr
$10$ & $2.984$ & $238(16)$  &  $218(17)$ & $254(20)$ \cr \hline
\end{tabular}
\label{tab:chicool}
\end{table}

The average charge $\langle q
\rangle _v (\tau)$ in volumes identified with instantons at $\tau=10$
is shown in Fig.~\ref{fig:tcc} versus the cooling time.  For the IA
action we find that $\langle q\rangle^{IA}_v (\tau)$ decreases
monotonically with $\tau$ which we interpret as (pair) annihilation of
small instantons during cooling. The behavior with the WA at small
$\tau$ is not clear, but for $\tau > 0.3$ $\langle
q\rangle^{WA}_v (\tau)$ shows a similar decrease.

We use the correlation between the average charge inside the instanton volume
as measured by the plaquette and geometric methods to
define the renormalization constant to be $Z_Q(\tau) = 
\langle q\rangle_v^{plaq.} (\tau) / \langle q\rangle_v^{geom.} (\tau=10)$. 
For uncooled lattices we find $Z^{WA}_Q(\beta=6.0) = 0.158(13)$, and 
$Z^{IA}_Q(\beta\approx6.0) = 0.230(9)$.  On varying the search parameters between $0.8
\le q_{min} \le 1.0$ and $24 \le r_{max}^2 \le 35$, the $Z_Q$ 
ranges roughly between $0.14$ and $0.17$ (WA), and $0.21$ to
$0.24$ (IA), $i.e.$ it has a rather weak dependence on the definition
of instanton volumes.  Including this dependence as a systematic
uncertainty, we estimate the renormalization constants to be 
\begin{equation} 
Z^{WA}_Q = 0.16(2), \qquad\quad Z^{IA}_Q  = 0.22(2).
\label{eq:zq}
\end{equation}
Our estimate of $Z^{WA}_Q$  is consistent with the value $0.18(2)$ obtained by
Alles {\it et al.\/} \cite{DiGiacomo94} using a heating method.  The larger value of
$Z^{IA}_Q$ is consistent with the expectation that the IA is a better
approximation of the continuum theory (where $Z_Q = 1$).  We use this 
estimate of $Z_Q(\tau)$ to compile the data presented in Table~\ref{tab:chicool}.

Our cooling procedure does not guarantee that all physical instantons
survive. We assume that instantons that survive are physical and their
properties, as a function of cooling time, can be studied once their
location is known.  The data show that instantons identified on cooled
lattices can be tracked back to the uncooled lattices, where UV noise
prevented their identification.  Systematic errors due to the
uncertainty in the scale and the non-uniqueness of the search
algorithm have not been fully included in the rough estimates of $\rho
$, density, and $Z_Q(\tau)$ given above.

\begin{figure}[t]
\def\axisscale{0.7}
\input {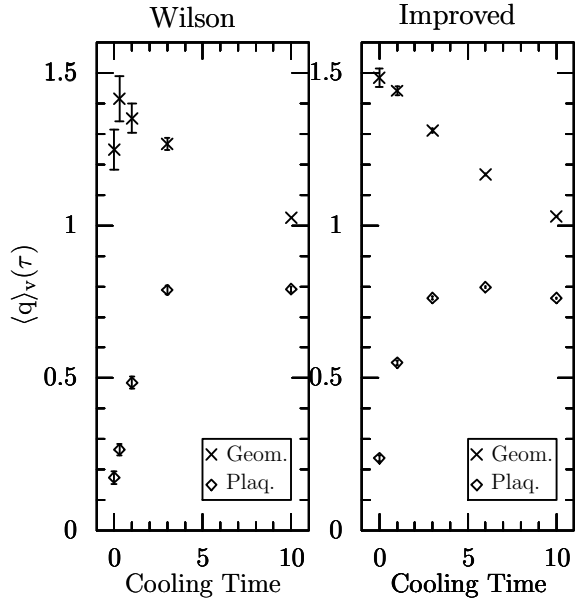}
\vspace{-28pt}
\caption{
Average charge $\langle q \rangle_v(\tau)$ 
contained within instanton volume versus $\tau$ 
for geometric (crosses) and plaquette (diamonds) methods.  }
\label{fig:tcc}
\end{figure}

%

We are refining this analysis in several ways. We are calculating the
string tension for the IA lattices to get a reliable estimate of the
scale. We are redoing the IA analysis with $\tau=30$ as the starting
point because the data in Fig.~\ref{fig:tcc} show that $\langle q
\rangle _v (\tau=10)$ with the geometric method has not stabilized.
We are investigating alternate schemes for identifying instanton
volumes, and are generating quark propagators to correlate hadronic
properties with instantons for the IA configurations.  Once the signal
in all these quantities is under control we plan to do the calculation
at weaker coupling to check scaling.

\end{document}